\documentclass[journal,comsoc]{IEEEtran}

\usepackage[T1]{fontenc}
\usepackage{url}

\usepackage{graphicx}
\usepackage{amsmath,amsthm}
\usepackage{mathtools}
\usepackage{subcaption}
\usepackage{color}
\usepackage{cite}
\usepackage{balance}
\usepackage[ruled,linesnumbered]{algorithm2e}

\begin{document}

\title{HASFC: a MANO-compliant Framework for \\ Availability Management of Service Chains}

\author{Mario~Di~Mauro,~
	Giovanni~Galatro, Maurizio~Longo,~ 
	Fabio~Postiglione, Marco~Tambasco
}

\maketitle

\begin{abstract}
Most softwarized telco services are conveniently framed as  Service Function Chains (SFCs). Indeed, being structured as a combination of interconnected nodes, service chains may suffer from the single point of failure problem, meaning that an individual node malfunctioning could compromise the whole chain operation.  
To guarantee ``highly available'' (HA) levels, service providers are required to introduce redundancy strategies to achieve specific availability demands, where cost constraints have to be taken into account as well. 
Along these lines we propose HASFC (standing for High Availability SFC), a framework designed to support, through a dedicated REST interface, the MANO infrastructure in deploying SFCs with an optimal availability/cost trade off. Our framework is equipped with: \textit{i)} an availability {\em model builder} aimed to construct probabilistic models of the SFC nodes in terms of failure/repair actions; \textit{ii)} a {\em chaining and selection} module to compose the possible redundant SFCs, and extract the best candidates thereof. 
Beyond providing architectural details, we demonstrate the functionalities of HASFC through a use case which considers the IP Multimedia Subsystem, an SFC-like structure adopted to manage multimedia contents within $4G/5G$ networks.         

\end{abstract}

\IEEEpeerreviewmaketitle

\section{Introduction and Motivation}
\label{sec:intro}

Pragmatically, High Availability (HA) is ``five nines" availability, i.e. the probability of $0.99999$ that a system is functioning when it is requested for use. Or else, the maximum tolerated yearly downtime amounts to $5$ minutes and about $15$ seconds. This is an industry benchmark historically adopted to characterize telecommunication systems (from traditional telephony to IP-based networks) in terms of their ability of providing a service. It also holds within modern networks, where the impelling {\em softwarization} process imposes new intriguing challenges \cite{mag1}. Virtualized and Containerized environments arising with the Network Function Virtualization (NFV) paradigm, in fact, drastically modify the structure of network nodes (e.g. router, switches, web servers, firewalls, and many others) whose software logic (often known as Virtual Network Function - VNF) is decoupled from the hardware resources through intermediate entities such as hypervisors or container engines. 

On one hand, such a decoupling has a beneficial effect in terms of scalability and flexibility, since more VNFs can share the same underlying infrastructure; on the other hand, the presence of ``nested'' layers (hardware, virtual machine, VNF) has the drawback of amplifying the so-called common mode failures. It means that an undesired malfunctioning of the virtual machine, for instance, causes the breakdown of all the VNFs sharing the same virtual machine. 

An additional issue to consider is the chained structure of nodes implementing the Service Function Chain (SFC) paradigm, which represents the modern way to provide services in NFV-based environments \cite{cerroni1,cerroni2}. A concatenation of nodes, in fact, directly implies that the malfunctioning of a single node may disrupt the whole service chain operation (single point of failure problem). 

Many network domains (e.g. radio access networks, data networks) embrace the service chaining concept, where the data flow traverses the virtualized network nodes in a predefined way. As an example of SFC-like structure we adopt the virtualized IP Multimedia Subsystem (vIMS), a crucial platform to manage multimedia communication within $4$G/$5$G domains, that we deploy onto a testbed relying on {\em Clearwater}. It is an open source deployment of IMS where all network nodes are virtualized.  

In general, when coping with softwarized environments, two critical capabilities should be considered (see e.g. \cite{tola}):
\begin{itemize}
	\item {\em Redundancy}: Replicate network elements or functionalities so as to manage the failures. In a chained structure such as the SFC, it is possible to replicate both the single layers (e.g. hardware, VNF) of a node, or the whole node belonging to the SFC, in order to preserve the continuity of the flow traversing the chain.
	\item {\em Predictive Maintenance}: Techniques aimed at estimating when the maintenance should be performed before fault conditions could provoke damages. 
\end{itemize}
The availability represents a pillar in the service continuity management and is strongly connected with the number of redundant nodes deployed in the service architecture and with the capability to maintain such nodes.
As a matter of fact, the availability requirements do not represent just a customer expectation, but obey to regulatory requirements of each country \cite{nfvava1}. 

From a practical point of view, the availability of a system (including an SFC infrastructure) is achieved through replication strategies, where costs for deploying the redundant elements must be taken into account.

\subsection{Steady-State Availability: a gentle definition}

\begin{figure}[t!]
	\centering
	\captionsetup{justification=centering}
	\includegraphics[scale=0.29,angle=90]{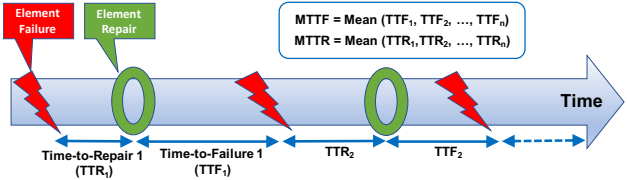}
	\caption{Mean-Time-to-Failure (MTTF) and Mean-Time-to-Repair (MTTR).}
	\label{fig:mttf}
\end{figure}

Concretely, one is interested at evaluating the so-called {\em steady-state} availability, namely the availability of a system during its regime condition. In other words, the steady-state availability is a constant quantity that can be evaluated when the system reaches a stable condition. Let us consider an element as the basic part of a complex system (e.g. a VNF as part of the whole SFC). By assuming a two-state model (up/down or, equivalently, working/failed) for a single element, two metrics are introduced to obtain the steady-state availability (see Fig. \ref{fig:mttf}): $i)$ Mean-Time-to-Failure (MTTF) which represents the average time interval between the restoration from a failed condition and a subsequent failure; $ii)$  Mean-Time-to-Repair (MTTR) which is the average time interval between the occurrence of a failure and the restoration activity. 

Being $A_v$ the steady-state availability, the following straightforward relationship holds:

\begin{equation}
 A_v= \frac{\textnormal{Uptime}}{\textnormal{Uptime~+~Downtime}} = \frac{\textnormal{MTTF}}{\textnormal{MTTF}+\textnormal{MTTR}}, 
 \nonumber
\end{equation}  
indicating that the steady-state availability is the ratio of the expected duration of a working state (MTTF) and the expected duration of a complete failure/repair cycle (MTTF~+~MTTR). When $A_v$ is greater than $0.99999$, we refer to the high (or five nines) steady-state availability. 
To obtain the steady-state availability of a complex system (made of elements variously interconnected among them as in the SFC case), we need to combine the steady-state availabilities of single elements. For the sake of simplicity, in the following we will use the term availability in place of steady-state availability.

\section{Related research and the offered Contributions}
\label{sec:rw}

Designing redundancy strategies to cope with the availability needs of service chains is brought to the attention of academic research and industrial development, but with some limitations in both fields. As regards the academic research, limitations include the adoption of methodologies useful to cope with the availability issues of softwarized networks, but scarcely scalable when the system becomes more complex (as in the case of the SFCs); it is the case of \cite{ctmc} where the Continuous-Time Markov Chain (CTMC) is adopted to model a virtualized node made of three nested layers: hardware, hypervisor, application layer. Likewise, the CTMC formalism has been adopted in \cite{dim_tsc} to encode a multi-state model of a VNF belonging to a service chain. One major drawback when applying Markov chains to model complex systems is the largeness of the pertinent state space. Conversely, in our framework we adopt the Stochastic Reward Networks (SRN), a technique allowing to compactly manage the space state of a real-world system and to take into account the probabilistic nature of failure/repair actions. Other works such as \cite{liu2016} are focused on the reliability evaluation of softwarized network chains, but take into account only failures and not repair actions; a repair modeling also lacks in \cite{fan2017}, where the focus is on the optimal number of VNFs to be deployed for guaranteeing a certain availability of the SFC. In contrast, we consider both of aspects, including correlated failures/repairs arising both from the nested layer of each virtualized nodes and from the chained interconnections among nodes themselves. Finally, there is no attempt (to the best of the authors knowledge) of designing a completely automated framework to manage the availability life-cycle of a complex structure as an SFC. In this line, our proposal appears to be completely original. 
The industrial development exhibits other kinds of limitations on its side. For example, the European Telecommunications Standards Institute (ETSI) released some generic guidelines for deploying service chains with redundancy mechanisms apt to satisfy a given availability demand \cite{etsi}. In particular, ETSI identifies the MANO (MANagement and Orchestration) as the principal component in charge of handling the availability of service chains through dedicated scaling/redundancy policies. Actually, such guidelines definitely lack methods or techniques useful to optimize the availability/cost trade off when deploying redundant service chains. Accordingly, many MANO projects and implementations are equipped with functionalities managing the VNFs lifecycle, but there is no record of a dedicated function logic to automatically govern the redundancy strategies in a structured way. Some examples include (see \cite{mano_proj1} for more detailed differences among the MANO projects): Open Baton, a platform which includes a Fault Management System based on alarms coming from the VIM; Open Network Automation Platform (ONAP) which includes a Drools-based policy module to provide a smart management of VNFs; Open Source Mano (OSM), a platform completely aligned with the ETSI NFV standard models, and exposing a rich set of REST APIs which allow to easily interact with external REST-based modules (such as the framework that we propose).

Going into the direction of embodying both academic and industrial aspects, we propose a framework which: $i)$ exploits robust and well-formalized techniques to deal with SFC availability issues supported by a dedicated algorithm, and $ii)$ can easily interact with the MANO, via the standard REST interface, to improve the whole SFC availability management.

\subsection{Contributions}

 This paper presents HASFC (which stands for High Availability SFC), a framework designed to automatically build service chains respecting a desired availability target at the minimum cost. The following main features can be recognized within the proposed framework:
\begin{itemize}
	\item {\em Model Building and Evaluation}: performed by a dedicated engine which allows to automatically obtain a probabilistic model of a virtualized node of an SFC in terms of its failure events and repair actions;  
	\item {\em Chain Composition and Selection}: a feature managed by a module performing two actions. The first one is to build the series of virtualized nodes belonging to the SFC along with the evaluation of the corresponding availability. The second one is to extract (through a customized and embedded heuristic search algorithm) a subset of SFCs satisfying the best trade off between availability requirements and deployment costs;  
	\item {\em REST API compliance}: although HASFC can be used as a stand alone framework, it exposes a REST interface, thus allowing the interaction with most of the existing MANO infrastructures. 
\end{itemize}
Most functionalities are embodied into a designed-from-scratch algorithm whose structure is inherited from the OptChains+ algorithm conceived by the same authors and available in \cite{dimtnsm}. 


\section{HASFC: a framework for high availability management of SFCs}
\label{sec:hasfc}

\begin{figure}[t!]
	\centering
	\captionsetup{justification=centering}
	\includegraphics[scale=0.25, angle=90]{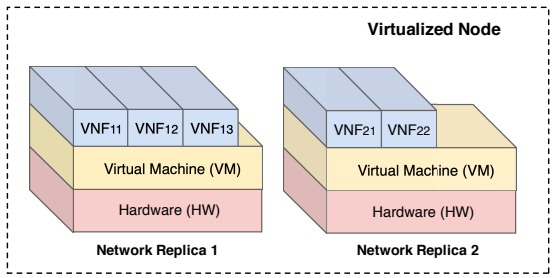}
	\caption{Virtualized SFC node composed of $2$ Network Replicas to improve the availability.}
	\label{fig:virtpcscf}
\end{figure}   

For redundancy and, then, for availability purposes, a virtualized node belonging to an SFC can be composed of one or more \textit{Network Replicas} (NR) as depicted in Fig. \ref{fig:virtpcscf}. The NR is a three-layered (the number of layers can be easily customizable) model including: $i)$ the Hardware (HW) layer, which represents the physical equipment (e.g. CPU, RAM, Power Supply); $ii)$ the Virtual Machine layer (VM) which is the interface allowing the decoupling between the underlying equipment and the software service logic; $iii)$ the VNF layer which includes a number of software instances providing a specific functionality. Inspecting the virtualized node model in Fig. \ref{fig:virtpcscf}, it is straightforward to recognize that a two-level redundancy can be implemented: by replicating single software instances belonging to the VNF layer, and/or by replicating the whole NR infrastructure. The VNF redundancy is useful to deal with the failure of one (or more) software instances. In contrast, the NR redundancy allows to face the common mode failures induced by malfunctioning of HW or VM causing the crash of all the software instances running onto an NR. Obviously, more redundant VNFs/NRs improve the availability of a virtualized node but entail higher deployment costs. Hence, HASFC automatically designs an optimal redundancy strategy to minimize the overall cost of VNFs/NRs replicas per node, by guaranteeing a given availability target (e.g. the five nines) for the whole SFC.

\begin{figure*}[t!]
	\centering
	\captionsetup{justification=centering}
	\includegraphics[scale=0.87]{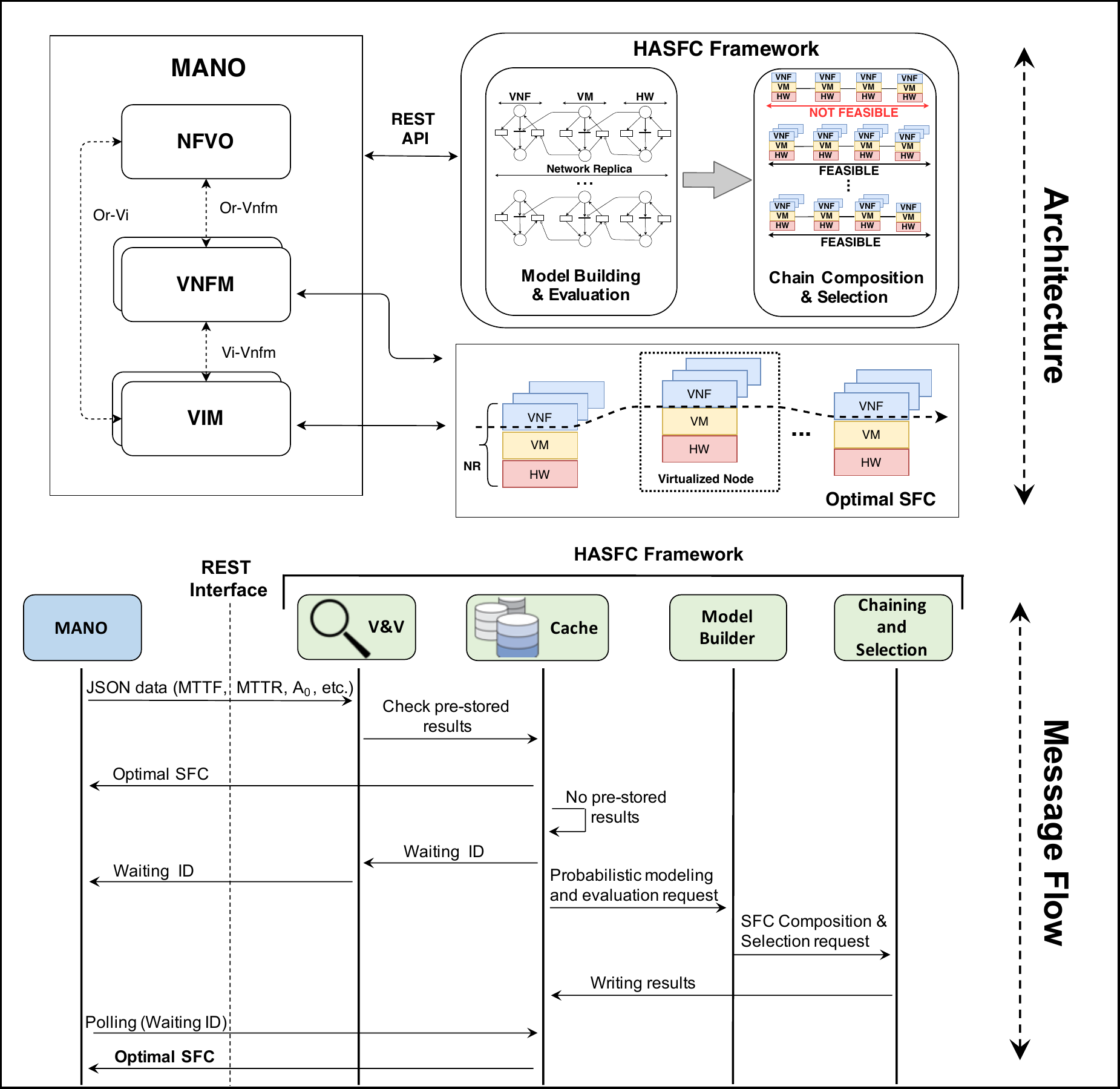}
	\caption{HASFC high-level architecture (topmost panel), and Message Flow (downmost panel).}
	\label{fig:hasfc}
\end{figure*}

\vspace{-0.08 cm}
\subsection{Architectural perspective of HASFC}
Figure \ref{fig:hasfc} (topmost panel) offers a high-level architectural perspective with MANO (the ``muscle'') and HASFC (the ``brain'') interacting to obtain an optimal SFC as output. Let us explore the two main functionalities composing the HASFC. 

The first one (Model Building and Evaluation) is designed to automatically build and evaluate the probabilistic model of an NR. Such a feature exploits the Stochastic Reward Networks (SRNs), a state-space based model \cite{trivedibook} which allows to characterize separately, in terms of failures and repairs, the three layers (VNF, VM, HW), and, then, combining them to form the whole NR. 
We recall that, with respect to other methods (e.g. CTMCs), SRN has the benefit of dealing with complex modeling scenarios since it provides a compact formalism to identify repetitive structures easily interpretable by non-experts, as well.
Precisely, the SRN technique allows to characterize each layer in terms of a binary condition (e.g. HW working / HW failed). Working and failure conditions are connected by means of two opposite transitions denoting two possible events: a failure event (working $\rightarrow$ failure, ruled by the MTTF) and a repair event (failure $\rightarrow$ working, ruled by the MTTR). Then, the three SRN models pertaining to VNFs, VM, and HW are connected by means of special transitions (called \textit{inhibitory arcs}) accounting for the nested layer structure. It means that, if the lower layer fails (HW), the remaining upper layers (VNFs and VM) are \textit{inhibited} to work; likewise, if the HW layer gets repaired, VNFs and VM layers get restored (as typically occurs when physical parts of a network architecture are repaired). Once built the multi-layered NR model, an internal routine relying on an SRN-based model tool (we adopt TimeNET \cite{timenet}, but other tools such as SHARPE \cite{trivedibook} can be exploited) allows to evaluate the availability of the considered NR, and, then, of the virtualized node.

The obtained availability value activates the second functionality of HASFC (Chain Composition and Selection) which performs two actions: \textit{i)} chaining the virtualized nodes to evaluate the availability of the resulting SFC; \textit{ii)} selecting the optimal set of SFCs maximizing the availability/costs trade off, through a dedicated routine performing an exhaustive search with pruning. 
It is useful to highlight that, since each virtualized node can be made of different NRs and each NR can host different VNFs, many combinations of SFCs can be obtained. To guarantee more flexibility, HASFC returns a number (customizable) of SFCs respecting the availability requirement at minimal cost. 

Once an optimal SFC is obtained, the MANO takes the responsibility of the SFC deployment where: \textit{i)} the NFV Orchestrator (NFVO) is in charge of the whole network service management, performing resource and network service orchestration; \textit{ii)} the VNF manager (VNFM) governs the VNFs lifecycle through operations such as instantiation, termination, scaling in/out (useful to increase/reduce the redundancy degree); \textit{iii)} the Virtualized Infrastructure Manager (VIM) is in charge of deploying the resulting SFC, by handling the association with physical resources (corresponding to lower layers of NRs). 
  
\subsection{Message flow within HASFC}

Let us explore the message flow between MANO and HASFC sketched in Fig. \ref{fig:hasfc} (downmost panel). The first message is a JSON-formatted request from MANO to HASFC. This message embodies input parameters (MTTF, MTTR, availability target A$_0$, etc.) used by HASFC to produce the desired optimal chain as output. The complete list of input parameters is provided below, where a practical use case is considered. 
Such a request arrives to the Verification and Validation (V\&V) module which performs a consistence/correctness check about the input data. Once passed the check, the request is transferred to a cache module which might contain pre-stored results in case identical requests have been made in the past. If this is the case, the cache module returns the optimal SFC to the MANO. Otherwise, the request is forwarded towards the Model Builder and Chain and Selection modules to evaluate the availability of the desired SFC and to pinpoint the optimal chain. 
In the meanwhile, MANO receives a message including a waiting ID to retrieve results once they are ready. Such results are written in cache, and MANO will query HASFC to get them. 
Depending on the specific request (e.g. type of SFC to build, number of NRs/VNFs needed, etc.), model building and chaining selection stages might be time consuming.
At this aim, we derive timing information for a specific use case involving the IP Multimedia Subsystem framework.

\begin{figure*}[t]
	\centering
	\captionsetup{justification=centering}
	\includegraphics[scale=0.48,angle=90]{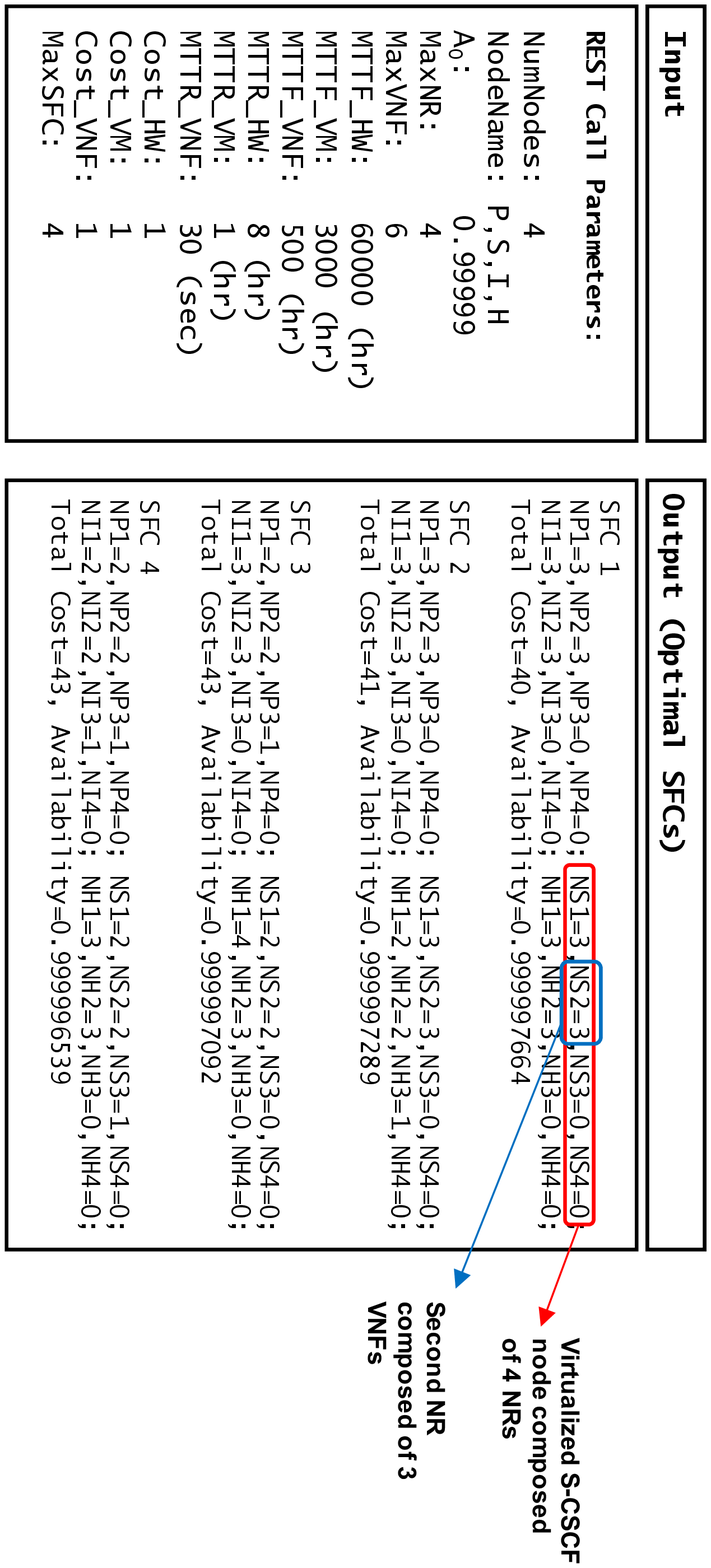}
	\caption{Optimal vIMS chain construction: REST input parameters (on the left), \\ and a list of resulting output SFCs (on the right).}
	\label{fig:output_sfc}
\end{figure*} 

\section{HASFC at work: Availability of virtualized IMS}
\label{sect:casestudy}

We analyze the working principles of HASFC by considering a virtualized IP Multimedia Subsystem (vIMS) as a valuable realization of an SFC. IMS plays a key role across a number of network technologies (including $5$G), providing support for evolved services (e.g. HD Voice/Video, online gaming, etc.). 
It is useful to recall that the vIMS exhibits a chained structure (see \cite{dim_tsc} for details) composed of $3$ virtualized Call Session Control Function (CSCF) nodes and a virtualized database server as detailed in the following: 

\textbf{Proxy CSCF (P-CSCF/P)}: it represents the first contact point with the IMS domain, aimed at managing the incoming requests (e.g. Invite, Register) received by subscribers. 

\textbf{Serving CSCF (S-CSCF/S)}: it is a critical node involved in controlling the status and the routing of an IMS session. 

\textbf{Interrogating CSCF (I-CSCF/I)}: it provides information about the S-CSCF to contact during an IMS session.

\textbf{Home Subscriber Server (HSS/H)}: it is an evolved database server where subscribers information (authentication keys, public/private identities, etc.) are stored.

\subsection{Optimal vIMS chain construction} 
 
We analyze how the MANO builds an optimal vIMS chain with the desired availability at the minimum cost, through the HASFC framework. Figure \ref{fig:output_sfc} reports: on the left, the key input parameters embedded in the REST call from MANO to HASFC; on the right, a list of output SFCs returned from HASFC to MANO.

The input parameters accepted by the REST call include: the number and name of nodes; the desired availability target A$_0$ (five nines here); the maximum number of network replicas (\textit{MaxNR}) to form a virtualized node; the maximum number of VNFs (\textit{MaxVNF}) to be hosted on top of a network replica; MTTF and MTTR values per NR layer (VNF, VM, HW); the cost per single layer; the maximum number of SFCs to be saved in the HASFC cache (\textit{MaxSFC}). 
Remarkably, even if the MANO will deploy only one SFC, we offer the flexibility of storing in cache more than one configuration through the \textit{MaxSFC} parameter. 

Once the input has been processed by internal modules of the HASFC as detailed in Sect. \ref{sec:hasfc}, the output appears as a list of four SFCs (\textit{MaxSFC} $=4$ in this use case) as reported in the right panel of Fig. \ref{fig:output_sfc}. Let us consider the first returned SFC, namely SFC $1$ (SFCs are sorted by the best availability/cost trade off), that will be considered by the MANO for the deployment. 
Each node is composed by max $4$ NRs (\textit{MaxNR} $=4$), and each NR is composed by max $6$ VNFs (\textit{MaxVNF} $=6$). For example, the S-CSCF node exhibits the configuration [NS1=3, NS2=3, NS3=0, NS4=0] denoting that it is composed of $2$ NRs with $3$ VNFs hosted on top, respectively. Although not present, we also include the third and fourth NRs having a null value. The final SFC exhibits an availability value amounting to $0.999997664$ at the minimal cost of $40$. To compute the total SFC cost, we consider separately the cost contributions for each layer. For example, according to the input cost parameters chosen in this use case (HW, VM, and VNF costs all amounting to $1$), the S-CSCF cost amounts to: $2$ (HW) + $2$ (VM)  + $6$ (VNFs) = $10$. Then, the total SFC cost is obtained by summing up each node cost. In this use case, the assumption of equally priced layers roughly takes into account the fact that software (VNF) parts have a cost comparable with the remaining parts due to an extra amount of licenses. If it is difficult to estimate cost layers, only the SFC availability result could be considered as a useful output.  
The remaining SFCs also satisfy the five nines requirement, but at a worse availability/cost trade off. 

In our experiment, the HASFC framework has been installed on a standard PC equipped with Intel Core $i5-7200U$@$2.50$ GHz (quadcore) and with $16$ GB of RAM (main code and installation guidelines are available in \cite{hasfc}). The time required to obtain the resulting SFCs highly depends on the number of nodes composing the SFC, the number of redundant NRs per node, and the number of VNFs to be hosted on each NR. The Model Building feature is the most time consuming. It builds a stochastic model for each NR, and the time spent to evaluate the single model ranges from a couple of minutes for $2$ VNFs up to $2$ hours for $6$ VNFs. Thus, an operator which needs a real-time deployment of an SFC must be aware of the time required to perform the availability evaluation.

In contrast, the Chaining and Selection functionality is less demanding. With the considered input parameters ($4$ nodes, \textit{MaxNR}$=4$, \textit{MaxVNF}$=6$), more than one hundred million SFCs are built. 
Such an intractable number is reduced to about $10,000$, thanks to the heuristic search algorithm with pruning embedded in the Chaining and Selection module. 
Such a stage requires approximately $3$ minutes. Finally, we perform additional load/stress tests including: the adoption of more complex NR structures (which include up to five layers as in many container-based architectures); sensitivity tests to evaluate the availability values oscillations for small deviations of some input parameters (e.g. MTTF, MTTR); automated REST calls to HASFC.

\subsection{vIMS chains comparison}

We now compare some resulting SFCs, imposing three different availability requirements: four nines, five nines, and six nines. Let us consider $9$ SFCs divided into three groups: the first group contains $3$ SFCs obtained with the four nines availability requirement ($S^{(4)}_1$, $S^{(4)}_2$, $S^{(4)}_3$). Similarly, the other two groups contain $3$ five-nines SFCs ($S^{(5)}_1$, $S^{(5)}_2$, $S^{(5)}_3$) and $3$ six-nines SFCs ($S^{(6)}_1$, $S^{(6)}_2$, $S^{(6)}_3$), respectively.  

\begin{figure}[t]
	\centering
	\captionsetup{justification=centering}
	\includegraphics[scale=0.37]{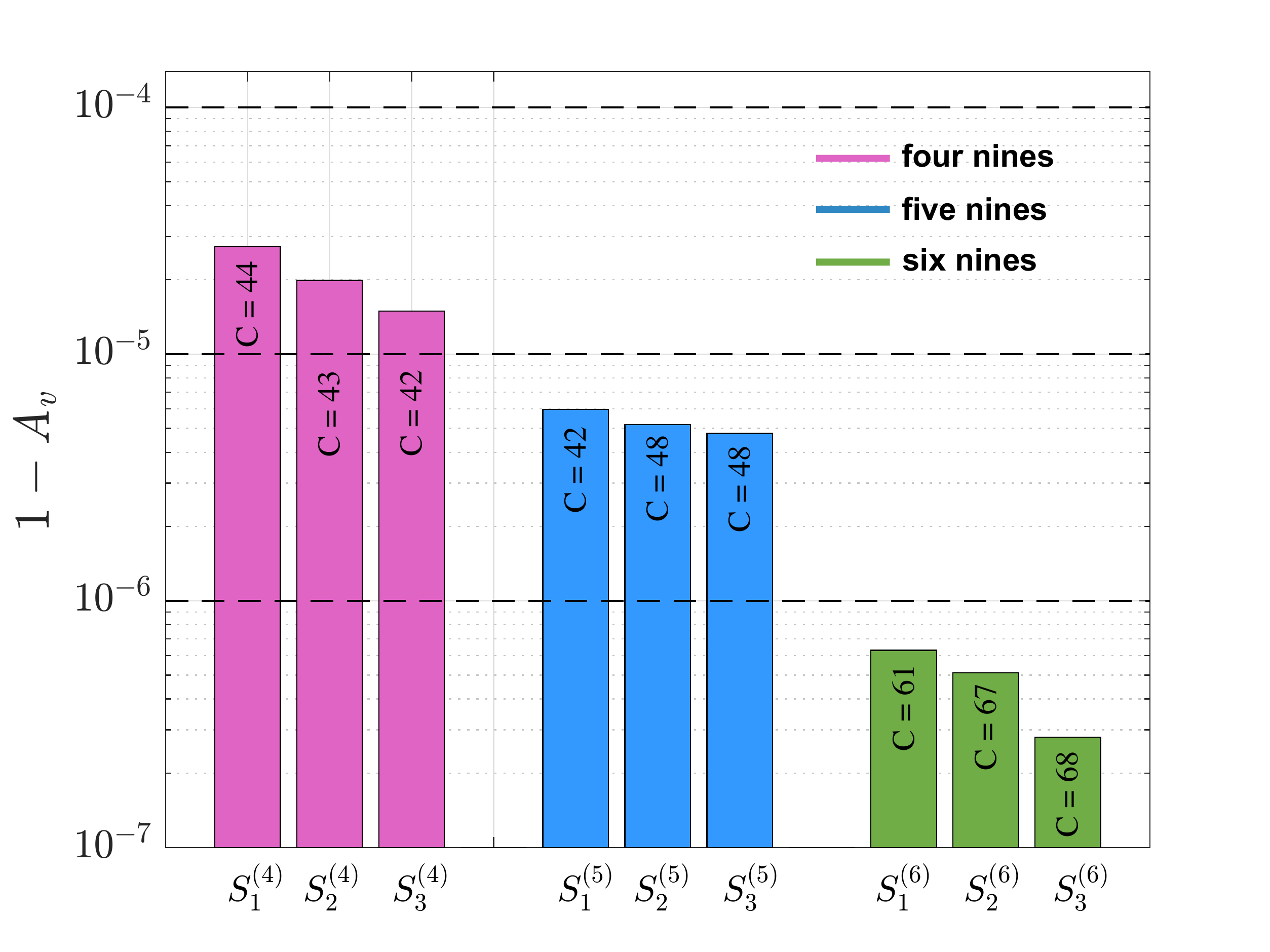}
	\caption{vIMS chains comparison.}
	\label{fig:bars}
\end{figure}

Such SFCs are compared in terms of availability/costs as shown in Fig. \ref{fig:bars}. The x-axis reports the three groups of SFCs identified by different colors (four nines/violet, five nines/blue, six nines/green). For ease of visualization, the y-axis reports the value of unavailability ($1-A_v$) associated to each SFC. Through such a representation, it is immediate to recognize the SFCs respecting the five nines condition (blue bars), since the corresponding unavailability values lie below the horizontal dashed line at $10^{-5}$. In contrast, to satisfy the extremely challenging six nines requirement (approximately $31$ seconds of maximum yearly tolerated downtime) the SFCs unavailability values must lie below the horizontal dashed line at $10^{-6}$ (green bars). Within each bar we report the total SFC cost $C$. 

Intuitively, SFCs with higher costs should exhibit higher availability values (or lower unavailability values) due to the higher redundancy degree. Such a condition can be violated if we compare, for example, $S^{(4)}_1$ and $S^{(5)}_1$. The former (clearly not optimal) exhibits a four-nines availability with cost $C=44$, whereas the latter exhibits a five-nines availability at a cost $C=42$. This apparently weird behavior is the consequence of different NR/VNF allocation policies across the virtualized nodes. Specifically, $S^{(4)}_1$ and $S^{(5)}_1$ exhibit the same configuration for P-CSCF and I-CSCF, having both nodes $2$ NRs and $3$ VNFs on top of each NR.
Moreover, for $S^{(4)}_1$ we have: the S-CSCF node with $4$ NRs and only one VNF on top of each NR, and the HSS node with $3$ NRs and $2$ VNFs on top of each NR. For $S^{(5)}_1$ we have: the S-CSCF node with $2$ NRs and $3$ VNFs on top of each NR, and the HSS node with $2$ NRs and $4$ VNFs on top of each NR. 
It appears clear that, in case of $S^{(5)}_1$ there are more VNFs for S-CSCF ($6$ VNFs) and HSS nodes ($8$ VNFs) w.r.t. $S^{(4)}_1$, but they are more concentrated on few NRs. Such an allocation policy guarantees more stability. As a result, $S^{(5)}_1$ is more robust and less expensive than $S^{(4)}_1$.   

Interestingly, to fulfill the six nines requirement, we have to involve all the available NRs with more than one VNF on top, resulting in an obvious increasing cost. For example, $S^{(6)}_1$ exhibits an availability value amounting to $0.99999937$ and a cost of $61$. 
In such a case, the behavior is quite expected: higher costs are associated to higher availability values. This is due to the fact that each node exploits nearly all the four available NRs. It implies a strong infrastructure redundancy, thus, six nines availability values are straightforwardly achieved. 
 
\vspace{-1mm}
 
\section{Conclusion and Research Prospects}

This work presents HASFC, a MANO-compliant platform for the availability management of service chains. Through a dedicated REST interface, HASFC empowers the basic MANO functionalities by receiving information about the desired chain to build, and returns a short list of SFCs satisfying availability/costs requirement. The whole framework embeds: $i)$ an engine able to build a probabilistic model of an SFC node through the powerful Stochastic Reward Network technique; $ii)$ a customized module to compose the whole SFC model and to select the chains with a given availability/cost trade off, through a heuristic with pruning search routine.    
The proposed framework can find useful applications in the field of softwarized network  management, and can be extended in several directions such as: $i)$ Including performance evaluation of SFCs taking into account additional metrics such as Quality-of-Service (QoS) or Quality-of-Experience (QoE); $ii)$ Considering more sophisticated chained architectures with additional nested layers; $iii)$ Embedding artificial intelligence mechanisms for smart resource management and automatic service provisioning in line with the impelling $6$G paradigm.

\balance
\vspace{200pt}
\begin{IEEEbiographynophoto}
	{Mario Di Mauro} (mdimauro@unisa.it) is a Research Fellow and Adjunct Professor with University of Salerno (Italy). His main fields of interest include network security, network availability, and statistical data analysis for telecommunications.
\end{IEEEbiographynophoto}

\vspace{-55pt}
\begin{IEEEbiographynophoto}{Giovanni Galatro} (galatrogiovanni@gmail.com) is a Cloud Engineer at Amazon AWS. His main fields of interest include network availability and machine learning applied to networking problems.
\end{IEEEbiographynophoto}

\vspace{-55pt}
\begin{IEEEbiographynophoto}
{Maurizio Longo} (longo@unisa.it) retired in 2018 from the University of Salerno (Italy)
as Full Professor of Telecommunications, having served as
Dean of Department and as Chairman of the Graduate School of Information
Engineering. He held academic positions also with the University Federico
II and the Parthenope University (Napoli), the University of Lecce and the
Aeronautical Academy. He has authored about 200 papers, mainly in the fields
of telecommunication networks and statistical signal processing.
\end{IEEEbiographynophoto}

\vspace{-55pt}
\begin{IEEEbiographynophoto} {Fabio Postiglione} (fpostiglione@unisa.it) is an Assistant Professor of Applied Statistics with the University of Salerno (Italy). His main research interests include statistics, availability and reliability of complex systems (e.g. telecommunication networks), degradation processes. He co-authored about 120 papers.
\end{IEEEbiographynophoto}

\vspace{-55pt}
\begin{IEEEbiographynophoto} {Marco Tambasco} (marco.tambasco@ericsson.com) is a senior Research Engineer at Ericsson Telecommunication Italy. His main fields of interest include availability and security of virtualized networks. 
\end{IEEEbiographynophoto}

\end{document}